\documentstyle[preprint,aps]{revtex}

\begin{document}

\draft

\title{A method for calculating the imaginary part of the Hadamard 
Elementary function $G^{(1)}$ in static, spherically symmetric 
spacetimes}

\author{Rhett Herman\cite{her}}
\address{Department of Chemistry and Physics, Radford University,
Radford, Virginia 24142}

\date{March 18, 1998}

\maketitle

\begin{abstract}
Whenever real particle production occurs in quantum field theory, the 
imaginary part of the Hadamard Elementary function $G^{(1)}$ is 
non-vanishing.  A method is presented whereby the imaginary part of 
$G^{(1)}$ may be calculated for a charged scalar field in a static 
spherically symmetric spacetime with arbitrary curvature coupling and 
a classical electromagnetic field $A^{\mu}$.  The calculations are 
performed in Euclidean space where the Hadamard Elementary function 
and the Euclidean Green function are related by ${1\over 
2}G^{(1)}=G_{E}$.  This method uses a $4^{th}$ order WKB approximation 
for the Euclideanized mode functions for the quantum field.  The mode 
sums and integrals that appear in the vacuum expectation values may be 
evaluated analytically by taking the large mass limit of the quantum 
field.  This results in an asymptotic expansion for $G^{(1)}$ in 
inverse powers of the mass $m$ of the quantum field.  Renormalization 
is achieved by subtracting off the terms in the expansion proportional 
to nonnegative powers of $m$ , leaving a finite remainder known as the 
``DeWitt-Schwinger approximation.''  The DeWitt-Schwinger 
approximation for $G^{(1)}$ presented here has terms proportional to 
both $m^{-1}$ and $m^{-2}$.  The term proportional to $m^{-2}$ will be 
shown to be identical to the expression obtained from the $m^{-2}$ 
term in the generalized DeWitt-Schwinger point-splitting expansion for 
$G^{(1)}$.  The new information obtained with this present method is 
the DeWitt-Schwinger approximation for the imaginary part of 
$G^{(1)}$, which is proportional to $m^{-1}$ in the DeWitt-Schwinger 
approximation for $G^{(1)}$ derived in this paper.

\end{abstract}

\pacs{}

\section{Introduction}

Despite the abscence of a full quantum theory of gravity, there are 
well known methods to calculate useful quantities involving quantum 
fields in curved space.  One of the most successful of these methods 
is the semiclassical approach, wherein the gravitational and 
electromagnetic fields are treated classically while the various 
matter fields that act as sources are treated quantum mechanically.  
In this approach, the evolution of the spacetime from a set of given 
initial conditions is described by the combined Einstein-Maxwell Field 
Equations,
\begin{equation}
 G_{\mu\nu}=8\pi\langle T_{\mu\nu}\rangle,
\label{efe}\end{equation}
and
\begin{equation}
 F^{\mu\nu}{}_{;\nu}=4\pi\langle j^{\mu} \rangle,
\label{mfe}\end{equation}
where $\langle T_{\mu\nu}\rangle$ and $\langle j^{\mu}\rangle$ are the 
vacuum expectation values (VEVs) of the stress-energy tensor and the 
current due to a charged quantized scalar field.  This paper follows 
the sign conventions of Misner, Thorne, and Wheeler \cite{mtw} and 
uses natural units ($G = c = \hbar =1$) throughout.

In the semiclassical regime, the calculation of quantities such as 
$\langle\phi^{2}\rangle$ and $\langle T_{\mu\nu}\rangle$ may begin 
with calculation of the Feynman Green function $G_{F}(x,x')$ defined 
by \cite{bd}
\begin{equation}
 iG_{F}(x,x')=<0\langle T[\phi(x)\phi(x')]\rangle 0>,
\label{gfdef}\end{equation}
where $T[\ldots]$ is the time ordering operator, $\phi(x)$ is the 
quantum field evaluated at the stationary spacetime point $x$, and 
$\phi(x')$ is the quantum field evaluated at the nearby point $x'$.  
One of the major difficulties in using the semiclassical method in 
quantum field calculations is the fact that infinities appear in the 
VEVs in Eqs.(\ref{efe})-(\ref{mfe}) in both flat 
\cite{Schweber,BD1,BD2} and curved \cite{bd,bsd} spacetimes.  Various 
methods have been used to isolate those infinities and remove them 
from the physical theory \cite{bd}, yet arguably the most powerful of 
these methods is based on the classic work of Schwinger 
\cite{schwinger}.

Schwinger calculated the Feynman Green function for a quantized 
fermion field by introducing a fictitious, non-quantum mechanical 
Hilbert space.  This $(4+1)$-dimensional Hilbert space was constructed 
from the $4$ spacetime dimensions in addition to a fifth dimension 
which was identified as the proper time parameter $s$ within this 
fictitious space.  Working within this proper time space, Schwinger 
was able to both isolate the divergences in the quantum field 
integrals for the fermion current $\langle j^{\mu}(x)\rangle = 
\lim_{x'\rightarrow x} ie\mbox{tr}[\gamma^{\mu}G_{F}(x,x')]$ produced 
by an external electromagnetic field in flat spacetime, and to use 
those divergences to renormalize the charge $e$ of the quantum field.

Schwinger's calculations of the quantum action functional $W$ were 
performed by transforming the integrals involving the $4$ spacetime 
dimensions into the momentum representation.  The integrals were 
evaluated using perturbation expansions in powers of $eA_{\mu}$ and 
$eF_{\mu\nu}$, where $A_{\mu}$ and $F_{\mu\nu}$ are the gauge field 
vector and the electromagnetic field tensor, respectively.  Were these 
integrals to be evaluated without further modification, Schwinger 
noted \cite{schwinger6} that conservation of energy and momentum would 
dictate that no pair creation would occur, or $\langle 
j^{\mu}\rangle=0$.  Schwinger's solution to this situation was to add 
an infinitesimal imaginary part to the action integral, resulting in 
$W$ acquiring a positive imaginary part.  Schwinger associated this 
imaginary part of $W$ with pair production by stating that
\begin{equation}
 \left|e^{iW}\right|^{2}=e^{-2Im[W]}
\label{nopair}\end{equation}
represented the probability that no pair creation would occur.  After 
this identification was made, Schwinger's calculation of the pair 
creation rate yielded a series expansion for the probability of pair 
creation per unit four volume $\Gamma$ by a constant external electric 
field,
\begin{equation}
 \Gamma={e^{2}\over 4\pi^{3}}E^{2}
 \sum_{n=1}^{\infty}n^{-2}e^{\left({-n\pi m^{2}\over eE}\right)},
\label{schwingerrate}\end{equation}
where $m$ and $e$ are the mass and charge of the fermion field, 
respectively, and $E$ is the magnitude of the electric field.  The 
expansion of Eq.(\ref{schwingerrate}) yields a series involving 
inverse powers of $m^{2}$.

Schwinger's calculation for the pair creation rate assumed that 
spacetime was flat.  Building on the work of Schwinger, DeWitt 
\cite{dtgf,bsd} extended the proper time method to include curved 
spacetime results.  Using what is now called the DeWitt-Schwinger 
proper time method, DeWitt calculated an asymptotic expansion for the 
Feynman Green function for a real scalar field in an arbitrary curved 
spacetime.  This asymptotic expansion was an expansion in inverse 
powers of $m^{2}$, where $m$ is the mass of the quantum field.  This 
expansion of $G_{F}(x,x')$ yielded an expression with both real and 
imaginary parts.  The real and imaginary parts of $G_{F}(x,x')$ could 
then be identified with other Green functions by using the identity
\begin{equation}
 G_{F}(x,x')=\overline{G}(x,x')-{1\over 2}iG^{(1)}(x,x'),
\label{gfg1relation}\end{equation}
where $\overline{G}(x,x')$ is one-half the sum of the advanced and 
retarded Green functions, and the Hadamard Elementary function 
$G^{(1)}(x,x')$ \cite{bd} is defined by
\begin{equation}
 G^{(1)}(x,x')\equiv \langle 0|\{\phi(x),\phi(x')\}|0\rangle.
\label{g1def}\end{equation}
Thus, the DeWitt-Schwinger asymptotic series expansion for 
$G^{(1)}(x,x')$ has the general form,
\begin{equation}
 G^{(1)}(x,x')\sim
 B_{+2}(x,x')m^{2}+B_{0}(x,x')m^{0}+B_{-2}(x,x')m^{-2}+\ldots\ ,
\label{g1morder}\end{equation}
where the $B_{n}$ are coefficients constructed from curvature tensors.

In the semiclassical method, quantities such as 
$\langle\phi^{2}\rangle$ and $\langle T_{\mu\nu}\rangle$ may be 
calculated from $G^{(1)}(x,x')$ by taking the appropriate derivatives 
with respect to either the stationary spacetime point $x$ or the 
nearby point $x'$ \cite{bsd,dtgf,chr,chrthesis,bd}; e.g.,
\begin{equation}
 \langle\phi^{2}\rangle=\lim_{x'\rightarrow x}{1\over 2}G^{(1)}(x,x'),
\label{divphi2}\end{equation}
and
\begin{equation}
 \langle{T}^{\mu\nu}(x)\rangle = \lim_{x'\to x}Re \Big[
   {1\over 2}({1\over 2}-\xi)
   (g^\mu{}_{\tau'}G^{(1)|\tau'\nu}+g^\nu{}_{\rho'}G^{(1)|\mu\rho'})
   +(\xi-{1\over 4})
   g^{\mu\nu}g^{\alpha\rho'}G^{(1)}{}_{|\alpha\rho'}\ldots\Big] .
\label{divstress}\end{equation}
In Eq.(\ref{divstress}), the vertical bar ``$|$'' indicates gauge 
covariant differentiation, and $g^{\mu}{}_{\tau'}$ is the bivector of 
parallel transport \cite{dtgf}, which is used to transport the 
information from the nearby point $x'$ to the stationary point $x$.  
The infinities appearing in the VEV in Eq.(\ref{divphi2}) lead to the 
well known divergences in the unrenormalized expression for 
$\langle{T}^{\mu\nu}(x)\rangle$ in Eq.(\ref{divstress}) \cite{bd}.  
DeWitt's asymptotic expansion of $G^{(1)}(x,x')$ in powers of $m^{2}$ 
isolated the infinities, or the ``counterterms,'' of both 
$G^{(1)}(x,x')$ and $\langle T_{\mu\nu}\rangle$, with the infinities 
being found as coefficients of the $m^{2}$ and $m^{0}$ terms in the 
series of Eq.(\ref{g1morder}) \cite{chr}.  Once these infinities have 
been subtracted from the unrenormalized VEVs in 
Eqs.(\ref{divphi2})-(\ref{divstress}), then the finite remainder 
carries the information about the physics of the spacetime.  While 
these subtractions may be performed in principle, the remaining 
expressions are often so complicated that they can not be evaluated 
analytically \cite{bd}.

Investigators have used the semiclassical method to calculate 
renormalized expressions for both the vacuum polarization 
$\langle\phi^{2}\rangle_{ren}$ \cite{pc,vpf1,vpf2,ch,cj,pra}, and the 
stress energy tensor $\langle T_{\mu\nu}\rangle_{ren}$ 
\cite{hch,fz,bop,ahs} in specific spacetimes.  In each of these cases, 
the point-splitting counterterms were used to renormalize the 
divergent VEVs.  These calculations were performed in spacetimes whose 
high degrees of symmetry allowed the subtractions of the counterterms 
to be evaluated analytically.  See Ref.\cite{ahs} for a more thorough 
discussion of the difficulties encountered in these subtractions.

The DeWitt-Schwinger point-splitting expansion for $G^{(1)}$ may also 
do more than simply isolating the divergent counterterms necessary to 
renormalize the VEVs encountered in semiclassical quantum field 
theory.  If the terms proportional to nonnegative powers of $m$ are 
discarded, the remaining terms in Eq.(\ref{g1morder}) constitute the 
``DeWitt-Schwinger approximation'' for $G^{(1)}$ 
\cite{ahs,rhthesis,rhwhunpub} for fields whose masses are large when 
compared to the magnitude of the coefficients $B_{n}$.  Retaining the 
first one or two of these terms has been shown to provide a close 
approximation to the exact results in several cases 
\cite{vpf2,fz,pra,ahs}.

In the case of charged scalar fields, $G^{(1)}$ involves the complex 
quantity
\begin{equation}
 G^{(1)}(x,x')\equiv \langle 0|\{\phi(x),\phi^{*}(x')\}|0\rangle,
\label{g1complexdef}\end{equation}
where $\phi(x)$ is the charged scalar field.  Even though $G^{(1)}$ is 
generally complex, since all of the coefficients $B_{n}$ are real, 
Eq.(\ref{g1morder}) only contains part of the information about 
$G^{(1)}$.  DeWitt has pointed out \cite{bsd} that the point-splitting 
asymptotic expansion of $G^{(1)}$ in inverse powers of $m^{2}$ is 
incapable of yielding the imaginary part of $G^{(1)}$.  Yet whenever 
real particle production occurs $G^{(1)}$ will have a non-zero 
imaginary part.  This is readily seen by examining the expression for 
the current due to a charged scalar field,
\begin{equation}
 j^{\mu}={1\over 4\pi}F^{\mu\nu}{}_{\nu}={ie\over 2}\Big[
  \{D^{\mu}\phi,\phi^{*}\}-\{D^{\mu}\phi,\phi^{*}\}^{*}\Big],
\label{current}\end{equation}
where $e$ is the charge of the field, $D^{\mu}\equiv 
(\nabla^{\mu}-ieA^{\mu})$ is the gauge covariant derivative, $A^{\mu}$ 
is the background gauge field, and the asterisk ${}^{*}$ denotes the 
complex conjugate.  Using Eq.(\ref{g1complexdef}), and making the 
transition from classical to quantum fields \cite{rhwh}, gives the 
vacuum expectation value of the current due to the coupling between 
the charged scalar field and the background gauge field,
\begin{equation}
\langle j^{\mu}(x)\rangle = \lim_{x'\to x}{ie\over 4}\left[
  \left( G^{(1)|\mu}+g^{\mu}{}_{\tau'}G^{(1)|\tau'} \right) -
  \left( G^{(1)|\mu}+g^{\mu}{}_{\tau'}G^{(1)|\tau'} \right)^{*}\right].
\label{divcurrent} \end{equation}
Since the coefficients $B_{n}$ in Eq.(\ref{g1morder}) are real, using 
the DeWitt-Schwinger point-splitting expansion involving only inverse 
powers of $m^{2}$ to construct the current $\langle j^{\mu}\rangle$ 
from Eq.(\ref{divcurrent}) can only yield a value of zero.  Yet the 
current is not identically zero, as evidenced by Schwinger's 
calculation in flat space, thus illustrating the limitation the 
DeWitt-Schwinger point-splitting expansion suffers by yielding only 
the imaginary part of $G^{(1)}$.

In this paper, a method is presented to obtain an analytic expression 
for the imaginary part of $G^{(1)}$ in a static spherically symmetric 
spacetime with a gauge field $A^{\mu}$.  The imaginary part arises in 
a straightforward manner due to the presence of the gauge field 
$A^{\mu}$ and does not require the addition of any extra terms during 
the course of the calculation.  The renormalized expression 
$G^{(1)}_{ren}$ would ordinarily be constructed by subtracting the 
DeWitt-Schwinger point-splitting counterterms from the unrenormalized 
expression for $G^{(1)}$;
\begin{equation}
 G^{(1)}_{ren}=G^{(1)}_{unren}-G^{(1)}_{DS}.
\label{g1ren}\end{equation}
The mode sums and integrals in Eq.(\ref{g1ren}) can not be evaluated 
analytically at present due to the complexity of the terms in the 
infinite sums and integrals.  A method recently described by Anderson, 
Hiscock, and Samuel (AHS) \cite{ahs} yields a way to construct an 
approximate expression for $G^{(1)}_{ren}$.  All subtractions are 
performed in Euclidean space where the relationships between various 
Green functions are given by \cite{pra}
\begin{equation}
 \langle\phi^{2}\rangle={1\over 2}G^{(1)}(x,x')
 =iG_{F}(x,x')=G_{E}(x,x').
\label{gfrelations}\end{equation}
Using the Euclideanized metric, the mode functions in the VEV of 
Eq.(\ref{g1complexdef}) may be rewritten in a WKB form \cite{pra,ahs}.  
The renormalized expression $G_{E,ren}$ is constructed by moving into 
Euclidean space and performing the subtraction
\begin{equation}
 G_{E,ren}=G_{E,WKB}-G_{E,WKBdiv},
\label{gewkbren}\end{equation}
where $G_{E,WKB}$ is the expression obtained in Euclidean space upon 
substituting the Euclidean WKB mode functions into 
Eq.(\ref{g1complexdef}), and $G_{E,WKBdiv}$ contains the ultraviolet 
divergences found in $G_{E,WKB}$.  The mode sums and integrals in 
Eq.(\ref{gewkbren}) still can not be evaluated analytically in their 
present form.  In order to construct an expression for $G_{E,ren}$, 
the fourth order WKB approximations for the exact mode functions are 
substituted into the quantum field expressions.  Then, the mode sums 
and integrals in the subtraction of Eq.(\ref{gewkbren}) may be 
evaluated analytically by expanding the integrands and summands in 
powers of the mass $m$ of the quantum field.  This mass is assumed to 
be large when compared to the inverse of the radius of curvature of 
the spacetime.  By working within this large mass limit, a 
DeWitt-Schwinger asymptotic expansion results for $G_{E,ren}$ similar 
to Eq.({\ref{g1morder});
\begin{eqnarray}
 G_{E,ren}&=&(G_{E,WKB}-G_{E,WKBdiv})_{large\ m}
 \\
 &\sim&B_{+2}(x,x')m^{2}+B_{+1}(x,x')m^{1}+B_{0}(x,x')m^{0}\\
 &&+B_{-1}(x,x')m^{-1}+B_{-2}(x,x')m^{-2}+\ldots\ .
\label{g1wkbmorder}\end{eqnarray}
Eq.(\ref{g1wkbmorder}) differs from Eq.(\ref{g1morder}) in that the 
presence of the gauge field in this method is directly responsible for 
the presence of the new terms involving odd powers of $m$.  
Renormalization is achieved by discarding terms proportional to 
nonnegative powers of $m$, leaving a finite DeWitt-Schwinger 
approximation for $G_{E,ren}$;
\begin{eqnarray}
 G_{E,ren}&\approx&B_{-1}(x,x')m^{-1}+B_{-2}(x,x')m^{-2}.
\label{g1wkbds}\end{eqnarray}
In Euclidean space, all of these terms are real.  When the final 
expression is rotated back to the Lorentzian sector, the real and 
imaginary parts of $G^{(1)}$ are obtained.  The terms proportional to 
$m^{-1}$ are all imaginary and are proportional to both odd powers of 
the gauge field $A^{\mu}$ and its derivatives along with odd powers of 
the charge $e$ of the quantum field.  The terms proportional to 
$m^{-2}$ are all real and are exactly those obtained from the $m^{-2}$ 
term of the DeWitt-Schwinger point-splitting expansion 
\cite{chr,pra,rhthesis}.

In Sec.\ \ref{sec:unrenormalized}, an expression is derived for the 
unrenormalized Euclidean Green function $G_{E}$ in a static 
spherically symmetric spacetime.  In this paper, the field is assumed 
to be at zero temperature, although this method would extend to the 
analysis of fields at nonzero temperature \cite{pra}.  Sec.\ 
\ref{sec:renormalization} renormalizes the expression for $G_{E}$ 
derived in Sec.\ \ref{sec:unrenormalized}.  The renormalization 
subtractions are evaluated analytically by evaluating the mode sums 
and integrals in the unrenormalized expression in the limit that the 
mass of the quantum field is large when compared to the inverse of the 
radius of curvature of the spacetime.  The resulting DeWitt-Schwinger 
approximation will be shown to contain both the real and imaginary 
parts of $G^{(1)}$, with the main goal of this paper being obtaining 
an expression for the imaginary part.

Sec.\ \ref{sec:results} contains the results of the calculation of the 
asymptotic expansion of $G^{(1)}$.  The terms proportional to $m^{-2}$ 
will be shown to give the same results as the $m^{-2}$ term in the 
generalized DeWitt-Schwinger point-splitting expansion when the 
general expression is evaluated in a static spherically symmetric 
spacetime \cite{chr,pra}.  The terms proportional to $m^{-1}$ are all 
real in Euclidean space but are all imaginary when rotated back to 
Lorentzian space.  All of the $m^{-1}$ terms are proportional to odd 
powers of the charge $e$ of the complex scalar field, as well as odd 
powers of the gauge field $A^{\mu}$ and derivatives, and thus these 
terms all vanish when either the quantum field is uncharged or the 
spacetime is uncharged.  This connection to previous work will serve 
as a check on the validity of the method presented here.

\section{Unrenormalized Green functions for the charged scalar field 
in Euclidean space}
\label{sec:unrenormalized}

The goal of this paper is to derive an expression for the imaginary 
part of $G^{(1)}$ so that curved space phenomena involving real 
particle production due to the presence of an electromagnetic field 
may be investigated.  Thus it is useful to begin with the 
electromagnetic vector potential for a static spherically symmetric 
electric field like that of the Reissner-Nordstr\"{o}m spacetime;
\begin{equation}
 A^{\mu}=(A^{t}(r),0,0,0).
\label{vectorpotential}\end{equation}
The calculations will be simplified by moving to Euclidean space 
\cite{pra}.  The Euclidean metric for a static spherically symmetric 
spacetime is
\begin{equation}
 ds^{2}=f(r)d\tau^{2}+h(r)dr^{2}+r^{2}d\Omega^{2},
\label{eucmetric}\end{equation}
where $\tau\equiv -it$, $f(r)$ and $h(r)$ are the same functions as in 
the Lorentzian metric for a static spherically symmetric spacetime
\begin{equation}
 ds^{2}=-f(r)dt^{2}+h(r)dr^{2}+r^{2}d\Omega^{2},
\end{equation}
and $t$ is the 
coordinate whose derivative $(\partial/\partial t)$ is the timelike 
Killing vector in the Lorentzian sector \cite{pra}.  The Lorentzian 
and Euclidean expressions for the invariant $A_{\mu}A^{\mu}$ are given 
by
\begin{eqnarray}
 A_{\mu}A^{\mu}&=&g_{tt}A^{t}_{L}A^{t}_{L}=-f(r)A^{t}_{L}A^{t}_{L}
                = f(r)(iA^{t}_{L})(iA^{t}_{L})
 \nonumber\\
 &=&g_{\tau\tau}A^{\tau}_{E}A^{\tau}_{E}
  = f(r)A^{\tau}_{E}A^{\tau}_{E}, 
\label{Asquared}\end{eqnarray}
where the subscripts ``L'' and ``E'' refer to the Lorentzian and 
Euclidean sectors, respectively.  Eq.(\ref{Asquared}) indicates that
\begin{equation}
 A^{\tau}_{E}=iA^{t}_{L}.
\label{AtauAt}\end{equation}
This shows the relation between the gauge invariant derivative 
operators in both sectors is
\begin{eqnarray}
 g_{\mu\nu}D^{\mu}_{L}D^{\nu}_{L}&=&
  g_{tt}(\nabla^{t}-ieA^{t})(\nabla^{t}-ieA^{t})+\cdots \nonumber\\
&=&-f(r)(\nabla^{t}-ieA^{t})(\nabla^{t}-ieA^{t})+\cdots  \nonumber\\
&=& f(r)(i\nabla^{t}-ie(iA^{t}))(i\nabla^{t}-ie(iA^{t}))+\cdots  
   \nonumber\\ 
&=& f(r)(\nabla^{\tau}-ieA^{\tau})(\nabla^{\tau}-ieA^{\tau})+\cdots  
   \nonumber\\ 
&=&g_{\tau\tau}D^{\tau}D^{\tau}+\cdots \nonumber\\
&=&g_{\mu\nu}D^{\mu}_{E}D^{\nu}_{E} \,
\label{eucderiv}\end{eqnarray}
indicating 
\begin{equation}
 D^{\tau}\equiv \nabla^{\tau}-ieA^{\tau}
\end{equation}
for the Euclidean sector.  

An explicit form for the Euclidean Green function for fields at zero 
temperature is needed.  The case of zero temperature is chosen here 
for simplicity, and the derivation here follows those of Anderson 
\cite{pra} and AHS \cite{ahs}.  The action of the wave operator on the 
Euclidean Green function is given by
\begin{eqnarray}
 [g_{\mu\nu}D^{\mu}_{E}D^{\nu}_{E}-(m^{2}+\xi R)]G_{E}(x,x')&=&
 -g^{-1/2}\delta^{4}(x,x')  \nonumber\\
&=&-{\delta(\tau,\tau')\delta(r,r'),\delta(\Omega,\Omega')\over
   r^{2}(fh)^{1/2}}.
\label{eucgf} \end{eqnarray}
The delta function $\delta(\Omega,\Omega')$ may be expanded using the 
Legendre polynomials with the result
\begin{equation}
 \delta(\Omega,\Omega')={1\over 4\pi}
 \sum_{l=0}^{\infty}(2l+1)P_{l}[\cos(\gamma)],
\label{angulardelta}\end{equation}
where $\cos(\gamma)=\cos(\theta)\cos(\theta')+ 
\sin(\theta)\sin(\theta')\cos(\phi-\phi')$.  With the scalar field 
chosen to be at zero temperature, the delta function for the split in 
the $\tau-\tau'$ direction is given by
\begin{equation}
 \delta(\tau,\tau')={1\over 2\pi}
 \int_{-\infty}^{\infty}d\omega\ e^{i\omega(\tau-\tau')}.
\end{equation}
Thus, the Euclidean Green function for a field at zero temperature is
\begin{equation}
 G_{E}(x,x')={1\over 4\pi^{2}}
 \int_{0}^{\infty}d\omega\ \cos[\omega(\tau-\tau')]
 \sum_{l=0}^{\infty}(2l+1)P_{l}[\cos(\gamma)]
 S_{\omega l}(r,r'),
\label{gfe}\end{equation}
where $S_{\omega l}(r,r')$ is an unknown function of the coordinate $r$.

Applying the wave operator of Eq.(\ref{eucderiv}) to $G_{E}$ gives a 
differential equation for $S_{\omega l}(r,r')$,
\begin{eqnarray}
 {1\over h}{d^{2}S\over dr^{2}}&+&\left[
 {2\over rh}+{1\over 2fh}{df\over dr}-{1\over 2h^{2}}{dh\over dr}
 \right]{dS\over dr} \nonumber\\
&-&
 \left[{(\omega-eA_{\tau})^{2}\over f}+{l(l+1)\over r^{2}}+
       m^{2}+\xi R \right]S=-{\delta(r,r')\over r^{2}(fh)^{1/2}},
\label{Sdiffeq}\end{eqnarray}
where $m$ is the mass of the field, and $A_{\tau}$ is the only 
non-zero component of the Euclideanized vector potential.  $S_{\omega 
l}(r,r')$ is given by
\begin{equation}
 S_{\omega l}(r,r')\equiv 
 C_{\omega l}p_{\omega l}(r)q_{\omega l}(r'),
\label{Swldef}\end{equation}
where $C_{\omega l}$ is a normalization factor, and $p_{\omega l}(r)$ 
and $q_{\omega l}(r')$ are the solutions of the homogeneous form of 
Eq.(\ref{Sdiffeq}).  In Refs.\cite{pra} and \cite{ahs}, $S_{\omega 
l}(r,r')$ is written in the form
\begin{equation}
 S_{\omega l}(r,r')\equiv 
 C_{\omega l}p_{\omega l}(r_{<})q_{\omega l}(r_{>}).
\label{paulspq}\end{equation}
$S_{\omega l}(r,r')$ satisfies the Wronskian 
condition
\begin{equation}
 C_{\omega l}\left[p_{\omega l}{dq_{\omega l}\over dr}-
 q_{\omega l}{dp_{\omega l}\over dr}\right]=
 -{1\over r^{2}}\left({h\over f}\right)^{1/2},
\label{wronskian}\end{equation}
obtained by integrating Eq.(\ref{Sdiffeq}) once with respect to $r$ 
from $r-\epsilon$ to $r+\epsilon$, with $\epsilon\rightarrow 0$ in the 
end.

The functions $p_{\omega l}$ and $q_{\omega l}$ may be put into a WKB 
form using \cite{pra}
\begin{eqnarray}
 p_{\omega l}(r)&=&{1\over \sqrt{2r^{2}W(r)}}
 e^{ \int^{r} W(r')\left({h\over f}\right)^{1/2}dr'}   \label{pwkb}\\
 q_{\omega l}(r)&=&{1\over \sqrt{2r^{2}W(r)}}
 e^{-\int^{r} W(r')\left({h\over f}\right)^{1/2}dr'}\ .\label{qwkb}
\end{eqnarray}
In Eq.(\ref{pwkb}), $p_{\omega l}(r)$ is well behaved at $r=0$ and the 
event horizons, but is divergent at $r=\infty$.  The function 
$q_{\omega l}(r)$ is divergent at $r=0$ and the event horizons, but is 
well behaved at $r=\infty$.  See Anderson \cite{pra} for a more 
complete discussion of the properties of these functions.  
Eqs.(\ref{wronskian})--(\ref{qwkb}) show that $C_{\omega l}=1$.  Using 
Eqs.(\ref{pwkb}) and (\ref{qwkb}), Eq.(\ref{Sdiffeq}) assumes the 
familiar WKB form
\begin{eqnarray}
 W^{2}&=&\Omega^{2}(r)+V_{1}(r)+V_{2}(r)+ \nonumber\\
&&
 {1\over 2}\left[
 {f\over hW}{d^{2}W\over dr^{2}}+
 \left({1\over h}{df\over dr}-{1\over fh^{2}}{dh\over dr}\right)
 {1\over 2W}{dW\over dr}-
 {3\over 2}{f\over h}\left({1\over W}{dW\over dr}\right)^{2}
 \right],
\label{wkb}\end{eqnarray}
where
\begin{eqnarray}
 \Omega^{2}(r)&\equiv& 
  (\omega-eA_{\tau})^{2}+m^{2}f+(l+{1\over 2})^{2}{f\over r^{2}},
  \label{Omdef}\\
 V_{1}(r)&\equiv&
  {1\over 2rh}{df\over dr}-{f\over 2rh^{2}}{dh\over dr}-
  {f\over 4r^{2}}, \label{V1def}\\
 V_{2}(r)&\equiv&
  \xi Rf=-\xi f \times \nonumber\\
 &&\left[
  {1\over fh}{d^{2}f\over dr^{2}}-
  {1\over 2f^{2}h}\left({df\over dr}\right)^{2}-
  {1\over 2fh^{2}}{df\over dr}{dh\over dr}+
  {2\over rfh}{df\over dr}-
  {2\over rh^{2}}{dh\over dr}+
  {2\over r^{2}h}-{2\over r^{2}} \right], \nonumber\\
 && \label{V2def}
\end{eqnarray}
and where the product $l(l+1)$ has been factored into two pieces 
according to
\begin{equation}
 l(l+1)=(l+{1\over 2})^{2}-{1\over 4}.
\label{llp1factor}\end{equation}

The simple form of the vector potential in Eq.(\ref{vectorpotential}) 
causes the WKB equation to assume a new but straightforward form.  
Previous work had WKB equations just like 
Eqs.(\ref{wkb})-(\ref{V2def}) with $A_{\tau}=0$ (see the references in 
Ref.\cite{ahs} for a more detailed list of work by others).  The new 
feature in the WKB differential equations is the presence of the 
electromagnetic vector potential.  Eq.(\ref{wkb}) indicates how the 
complex quantities necessary for determining the charged scalar field 
current $\langle j^{\mu}\rangle$ arise in the present method.  The WKB 
function $W(r)$, while real in the Euclidean sector, is now complex in 
Lorentzian space due to the electromagnetic field of 
Eq.(\ref{AtauAt}).

Eq.(\ref{wkb}) is not in general exactly solvable.  Fortunately, it 
may be solved approximately by iteration.  For example, the zeroth 
order solution for $W$ is $W=\Omega$.  Substituting 
$W\rightarrow\Omega$ into the right hand side of Eq.(\ref{wkb}) and 
solving for $W$ yields the second order WKB solution $W^{(2)}$,
\begin{eqnarray}
 W^{(2)}&=&\Omega+{1\over 2\Omega}(V_{1}+V_{2})
 +{1\over 4\Omega^{2}}V_{1}^{2}  \nonumber\\
&&
 +{1\over 4}\left[
 {f\over h\Omega^{2}}{d^{2}\Omega\over dr^{2}}+
 \left({1\over h}{df\over dr}-{f\over h^{2}}{dh\over dr}\right)
   {1\over 2\Omega^{2}}{d\Omega\over dr}-
 {3\over 2}{f\over h}{1\over \Omega^{3}}
   \left({d\Omega\over dr}\right)^{2}   \right].
\label{wkb2}\end{eqnarray}
The fourth order solution, $W^{(4)}$, is obtained by making the 
substitution $W\rightarrow W^{(2)}$ in the right hand side of 
Eq.(\ref{wkb}) and solving for $W$.  Just as in point-splitting, this 
iteration procedure could continue indefinitely.  However, in the 
present case of obtaining the $\vartheta(m^{-1})$ and 
$\vartheta(m^{-2})$ terms in the DeWitt-Schwinger expansion for 
$G^{(1)}$, the iterations need only be continued until the fourth 
order WKB solution, $W^{(4)}$, is reached.  As pointed out by AHS 
\cite{ahs}, a second order WKB expansion contains the information of 
the $\vartheta(m^{2})$ and $\vartheta(m^{0})$ terms in the 
DeWitt-Schwinger point-splitting expansion, while a fourth order WKB 
expansion contains the information of the $\vartheta(m^{-2})$ terms.  
Thus, the infinite counterterms in the theory may be reproduced by 
performing the present calculation using only $W^{(2)}$, while the 
DeWitt-Schwinger approximations of the $\vartheta(m^{-1})$ and 
$\vartheta(m^{-2})$ terms require the use of $W^{(4)}$.

For purposes of simplicity and to make connections to previous work 
\cite{ahs}, the choice is made to split the points along the 
$\tau-\tau'$ direction such that 
$(x,x')\rightarrow(\tau,r,\theta,\phi;\tau',r,\theta,\phi)$.  Making 
the definition \cite{ahs}
\begin{equation}
 A_{1}=\lim_{r'\rightarrow r}
 \sum_{l=0}^{\infty}\left[
 2(l+{1\over 2})p_{\omega l}(r)q_{\omega l}(r') \right]
 =\sum_{l=0}^{\infty}\left[{(l+{1\over 2})\over r^{2}W} \right],
\label{A1def}\end{equation}
with the limit $r'\rightarrow r$ being taken in the last expression, 
allows the unrenormalized Euclidean Green function to be written as
\begin{equation}
 G_{E,unren}(x,\tau;x,\tau')=\langle\phi^{2}\rangle_{unren}=
 {1\over 4\pi^{2}}\int_{0}^{\infty}
 d\omega\ \cos[\omega(\tau-\tau')]A_{1},
\label{gedef}\end{equation}
for the separation along $\tau-\tau'$.  The next section discusses the 
identification and removal of the two types of divergences that appear 
in this unrenormalized expression for $G_{E}(x,\tau;x,\tau')$.

\section{Renormalization of the $4^{th}$ order WKB approximation in the 
large mass limit}
\label{sec:renormalization}

There are two types of divergences which appear in Eq.(\ref{gedef}).  
The first of these is a superficial divergence which appears when 
evaluating the sum over $l$ with an upper limit of $l=\infty$, while 
the second is an ultraviolet divergence due to performing the 
integrations over $\omega$ with an upper limit of $\omega=\infty$.  
This section presents the details of how the subtractions of these 
divergent terms from the unrenormalized Green function may be 
performed and how the resulting expression may be evaluated 
analytically.

As discussed more fully elsewhere \cite{ahs,pra,ch,hch}, the 
superficial divergence due to the sums over $l$ can not be real since 
$G_{E,unren}$ must remain finite with the points separated as they are 
in Eq.(\ref{gedef}).  The sum and integral present in Eq.(\ref{gedef}) 
will be evaluated later by assuming the mass $m$ of the quantum field 
is large enough to expand the summand of $A_{1}$ in inverse powers of 
$m$.  At present, it is useful to consider how the false 
divergence over $l$ is isolated assuming the sum and integral of 
Eq.(\ref{gedef}) were to be evaluated without such a large mass expansion.

The superficial divergence that appears in $A_{1}$ may be identified 
by evaluating the sum
\begin{equation}
 \sum_{l=0}^{\infty} 2(l+{1\over 2})p_{\omega l}q_{\omega l}=
 \sum_{l=0}^{\infty}{(l+{1\over 2})\over r^{2}W}
\label{A1div}\end{equation}
in the limit $l\rightarrow\infty$.  The only part of $W$ that 
contributes as $l\rightarrow\infty$ is the zeroth order term of $W$, 
or $\Omega$; terms in $W$ proportional to negative powers of $\Omega$ 
do not diverge as $l\rightarrow\infty$.  Substituting $W=\Omega$ into 
$(l+{1\over 2})/r^{2}W$ and expanding in inverse powers of $l$, and 
then truncating the expansion at $\vartheta(l^{0})$ gives
\begin{equation}
 \sum_{l=0}^{\infty}\lim_{l\rightarrow\infty}{2(l+{1\over 2})
 \over 2r^{2}W}=\sum_{l=0}^{\infty}{1\over rf^{1\over 2}}.
\end{equation}
Thus, the counterterm $1/(rf^{1\over 2})$ must be subtracted from the 
summand of Eq.(\ref{A1def}) in order to remove the superficial 
divergence over $l$.  This now indicates the unrenormalized expression 
for $G_{E}(x,\tau;x,\tau')$ is given by
\begin{equation}
G_{E,unren}(x,\tau;x,\tau')=
 {1\over 4\pi^{2}}\int_{0}^{\infty}
 d\omega\ \cos[\omega(\tau-\tau')]\sum_{l=0}^{\infty}\left[
 {(l+{1\over 2})\over r^{2}W}-{1\over rf^{1\over 2}} \right].
\label{gelrenormalized}\end{equation}
This technique will serve to remove any superficial divergences over 
$l$ that may appear.  In Eq.(\ref{gelrenormalized}), only the term 
$1/(rf^{1\over 2})$ is subtracted from the single summand $(l+{1\over 
2})/r^{2}W$.  Later, when the WKB approximation for $W$ is substituted 
into Eq.(\ref{gedef}) and the summands and integrands are expanded in 
the large mass limit, many terms involving factors of $l$ will arise.  
Only some of those terms will contain a divergence in $l$, and those 
that do must have the appropriate counterterms in $l$ subtracted from 
them to remove this superficial divergence.

Eq.(\ref{gedef}) also contains the ultraviolet divergences that are 
known to arise in semiclassical quantum field theory.  These 
divergences occur as $\omega\rightarrow\infty$ and must be subtracted 
from the unrenormalized expression.  As in the case of the divergence 
over $l$, these subtractions must be performed term by term after the 
large mass expansions of $A_{1}$ have been performed.  Thus, the 
generalized form for the renormalized expression for $G_{E}$ is given 
by
\begin{eqnarray} 
 G_{E,ren}&=&
 {1\over 4\pi^{2}}\int_{0}^{\infty}
 d\omega\ \cos[\omega(\tau-\tau')]\times \nonumber\\
 &&\left[
 \sum_{l=0}^{\infty}
 \left\{{(l+{1\over 2})\over r^{2}W}\right\}_{WKB\ expansion\ terms}
 -\left(l\rightarrow\infty\ terms\right)
 -\left(\omega\rightarrow\infty\ terms\right)
 \right],
\label{gerenormalization}\end{eqnarray}
where the subscript ``WKB expansion terms'' indicates all of the terms 
that result from a $4^{th}$ order WKB expansion of $(l+{1\over 
2})/r^{2}W$, and the terms identified as divergent in $l$ and $\omega$ 
are subtracted from each of these terms of the expansion.

The calculation of all the terms in Eq.(\ref{gerenormalization}) is a 
straightforward but tedious process.  The general procedure follows 
Appendices D, F, and G in the recent work of AHS \cite{ahs} and will 
be described here.  The presence of the gauge field increases the 
difficulty of the calculations substantially, but the method of AHS 
can be modified to account for the gauge field.

The calculation of $G_{E,ren}$ starts with the substitution of the 
$4^{th}$ order WKB approximation for $W$ into the definition of 
$A_{1}$ and the subsequent identification of the various 
$l\rightarrow\infty$ counterterms.  To keep track of the order in the 
WKB expansions of quantities it is useful to introduce the 
dimensionless parameter $\alpha$, letting $\alpha\rightarrow 1$ in the 
end.  In Eq.(\ref{wkb}), $\Omega$ is $\vartheta(\alpha^{0})$, 
$V_{1}(r)$ is $\vartheta(\alpha^{1})$, and the rest of the terms are 
$\vartheta(\alpha^{2})$.  The quantity $(l+{1\over 2})/(r^{2}W)$ is 
expanded in powers of the WKB parameter $\alpha$, truncating the 
expansions at $\vartheta(\alpha^{4})$, and setting $\alpha=1$.  This 
will result in over 600 terms whose general form is given by
\begin{equation}
 L_{hjk}=\sum_{l=0}^{\infty}
 \left[{\omega^{h}2(l+{1\over 2})^{1+2j}\over \Omega^{k}}-
 (l\rightarrow\infty\ terms)\right],
\label{lhjkdef}\end{equation}
where $\Omega(r)$ is defined in Eq.(\ref{Omdef}), and various factors 
of $f(r), V_{1}(r),eA_{\tau}(r),\ldots,$ that are present in front of 
each of the $L_{hjk}$ have been omitted.  This definition of $L_{hjk}$ 
is similar to that defined as $L_{jk}$ in Ref.\cite{ahs}.  In the 
present work, the presence of the gauge field has led to the need to 
keep track of additional factors of $\omega$ which now enter into the 
definition of the $L_{hjk}$.  The notation here is chosen so that it 
most closely matches the notation of AHS. Determining the 
$l\rightarrow\infty$ subtraction terms proceeds as before by expanding 
the summands in each $L_{hjk}$ in invervse powers of $l$ and 
truncating the expansion at $\vartheta(l^{0})$.  For example, the 
expression for $L_{001}$ requires one subtraction term;
\begin{equation}
 L_{001}=\sum_{l=0}^{\infty}
 \left[{2(l+{1\over 2})\over \Omega}-{2r\over f^{1\over 2}}
 \right].
\label{L001}\end{equation}

Determining the $\omega\rightarrow\infty$ subtraction terms requires 
using the Plana sum formula \cite{ahs,ww}
\begin{equation}
 \sum_{l=k}^{\infty}g(l)=
 {1\over 2}g(k)+\int_{k}^{\infty}g(\tau)d\tau+
 i\int_{0}^{\infty}{dt\over e^{2\pi t}-1}\left[
 g(k+it)-g(k-it)\right].
\label{plana}\end{equation}
The first two terms of Eq.(\ref{plana}) may be computed analytically, 
while the third term is not in general able to be computed 
analytically.  Since the ultraviolet, or $\omega\rightarrow\infty$, 
behavior of these sums is dominant, the third term of Eq.(\ref{plana}) 
may be expanded in inverse powers of $\omega$.  The expansions are 
truncated at $\vartheta(\omega^{-1})$ since terms of 
$\vartheta(\omega^{-3})$ are not ultraviolet divergent \cite{ahs}.  
Then, the integrals that arise from the expansion in inverse powers of 
$\omega$ may be computed analytically.  Applying this procedure to 
$L_{001}$, for example, gives
\begin{eqnarray}
 L_{001}&=&
  \sum_{l=0}^{\infty}\left[
  {2(l+{1\over 2})\over \Omega}-{2r\over f^{1\over 2}}
  \right] \nonumber\\
 &=&{1\over 
   2\left[(\omega-eA_{\tau})^{2}+m^{2}f+{f\over 4r^{2}}\right]^{1/2}}
   \nonumber\\
 &&-{2r^{2}\over f}
   \left[(\omega-eA_{\tau})^{2}+m^{2}f+{f\over 4r^{2}}\right]^{1/2}
  \nonumber\\
 &&-{4\over \omega}\int_{0}^{\infty}dt{t\over e^{2\pi t}-1}\ +\cdots 
  \nonumber\\
 &=&
  -{2r^{2}\over f}\omega
  +{2er^{2}A_{\tau}\over f}
  +\left({1\over 12}-m^{2}r^{2}\right){1\over \omega}
  +\vartheta(\omega^{-3})\ .
\label{l001largew}\end{eqnarray}
The only $L_{hjk}$ for which there will be $\omega$ subtraction terms 
are those for which $\{j,k\}$ have the values 
$\{0,1\},\{0,3\},\{1,3\},\{1,5\},\{2,5\}$ and $\{2,7\}$; only $h=0$ 
occurs for these combinations of $\{j,k\}$.  Comparison of $L_{001}$ 
in this example with $L_{01}$ of AHS, along with Eq.(\ref{AtauAt}), 
show how the gauge field will bring complex quantities into the 
expressions for the $L_{hjk}$ and eventually into $G^{(1)}$.

Once the terms divergent in both $l$ and $\omega$ have been determined 
for each of the $L_{hjk}$, the expression for $G_{E,ren}$ is now given 
by
\begin{equation}
 G_{E,ren}=
 {1\over 4\pi^{2}}\int_{0}^{\infty}
 d\omega\ \cos[\omega(\tau-\tau')] 
 \{F(r)(L_{hjk}-L_{hjk,\omega\rightarrow\infty})\},
\label{gelwrenormalized}\end{equation}
where each of the $F(r)$ above represent the functions such as $f(r), 
V_{1}(r),eA_{\tau}(r),\ldots,$ that appear in the calculation but are 
not affected by the summations and integrations, and the curly 
brackets $\{\ldots\}$ indicate the large number of terms arising due 
to the WKB expansions.  In Eq.(\ref{gelwrenormalized}), the 
$\omega\rightarrow\infty$ counterterms for each $L_{hjk}$ are 
indicated symbolically by $L_{hjk,\omega\rightarrow\infty}$.  Since 
the terms which diverge as $\omega\rightarrow\infty$ are subtracted in 
Eq.(\ref{gelwrenormalized}), then the function 
$\cos[\omega(\tau-\tau')]$ will be dominated by the infinitesimal 
separation $(\tau-\tau')$.  This indicates that 
$\cos[\omega(\tau-\tau')]\approx 1$ is valid.  For the purposes of 
completeness, it should be noted that future work involving both 
$\langle j^{\mu}\rangle$ and $\langle T_{\mu\nu}\rangle$ will involve 
more factors of $\omega$ than those that arise within the $L_{hjk}$.  
Thus, it is useful to make the definition
\begin{equation}
 S_{ihjk}={1\over 4\pi^{2}}\int_{0}^{\infty}d\omega\ \omega^{i}
 (L_{hjk}-L_{hjk,\omega\rightarrow\infty}),
\label{sihjkdef}\end{equation}
where the indices $ihjk$ have been chosen so the present work is more 
easily compared to that of AHS. In the calculation of the Green 
function in this paper, $i=0$ throughout; in AHS, $i=0$ for the Green 
function calculation while $i=0$ or $2$ when calculating $\langle 
T_{\mu\nu}\rangle$.  The sums and integrals in Eq.(\ref{sihjkdef}) 
still can not be evaluated analytically in general in their present 
form.  They may be put into a form that can be evaluated analytically 
by taking the large mass limit of the quantum field.

Each of the $S_{ihjk}$ may be computed in the large mass limit by 
first expanding each $w^{i}2(l+{1\over 2})^{1+2j}/\Omega^{k}$ term in 
the $L_{hjk}$ using the Plana sum formula of Eq.(\ref{plana}).  The 
first two terms of the Plana formula may be computed exactly.  Again, 
the third term may not be computed analytically, but once expanded in 
the large mass limit each piece of the expansion may be computed 
analytically.  The large mass limit of this third term is obtained by 
expanding the integrand of the third term in inverse powers of the 
large quantity $\sqrt{(\omega-eA_{\tau})^{2}+m^{2}f+(l+{1\over 
2})^{2}f/r^{2}}$.  Each integral may then be evaluated analytically.  
After the integrations, the entire expression for each $S_{ihjk}$ is 
expanded in inverse powers of $m$.  In Appendix G of Ref.\cite{ahs}, 
AHS give an example of this type of calculation when they calculate 
the large mass expansion of $S_{001}$.  Here, the large mass expansion 
of $S_{0001}$ will be shown for comparison.

The $l\rightarrow\infty$ limit for $L_{001}$ is found in 
Eq.(\ref{L001}), while the $\omega\rightarrow\infty$ subtraction term 
is found in Eq.(\ref{l001largew}).  This gives the expression to be 
evaluated in the large mass limit as
\begin{eqnarray}
 S_{0001}&=&{1\over 4\pi^{2}}\int_{0}^{\infty}d\omega\ \omega^{0} \times
 \Bigg\{   \nonumber\\ 
&&\sum_{l=0}^{\infty}\Bigg[
  \left({2(l+{1\over 2})\over
  [(\omega-eA_{\tau})^{2}+m^{2}f+(l+{1\over 2})^{2}{f\over r^{2}}]^{1/2}}
  -{2r\over f^{1\over 2}}\right)   \nonumber\\
&&\qquad -\left(
  -{2r^{2}\over f}\omega
  +\left({1\over 12}-m^{2}r^{2}
  \right){1\over \omega}+{2er^{2}A_{\tau}\over f}\right)
  \Bigg] \Bigg\}.
\label{s0003def}\end{eqnarray}
Using the Plana sum formula and expanding in inverse powers of the 
large quantity $\sqrt{(\omega-eA_{\tau})^{2}+m^{2}f+(l+{1\over 
2})^{2}f/r^{2}}$ yields
\begin{eqnarray}
 S_{0001}&=&{1\over 4\pi^{2}}\int_{0}^{\infty}d\omega\Bigg\{ 
  \left({1\over 2}\right) \left(
  {1\over
  [(\omega-eA_{\tau})^{2}+m^{2}f+(l+{1\over 2})^{2}{f\over r^{2}}]^{1/2}}-
 {2r\over f^{1\over 2}} \right)   \nonumber\\ 
 &&+\int_{0}^{\infty}dl
 \left({1\over
  [(\omega-eA_{\tau})^{2}+m^{2}f+(l+{1\over 2})^{2}{f\over r^{2}}]^{1/2}}-
 {2r\over f^{1\over 2}} \right)   \nonumber\\ 
 &&+i\int_{0}^{\infty}{dt\over e^{2\pi t}-1}\bigg\{
 {2(it+{1\over 2})\over 
  [(\omega-eA_{\tau})^{2}+m^{2}f+(it+{1\over 2})^{2}{f\over 
  r^{2}}]^{1/2}}  \nonumber\\
 &&\qquad-{2(-it+{1\over 2})\over 
  [(\omega-eA_{\tau})^{2}+m^{2}f+(-it+{1\over 2})^{2}{f\over 
  r^{2}}]^{1/2}} \bigg\} \nonumber\\
 &&-\left(
  -{2r^{2}\over f}\omega
  +\left({1\over 12}-m^{2}r^{2}
  \right){1\over \omega}+{2er^{2}A_{\tau}\over f}\right)\Bigg\},
\end{eqnarray}
\begin{eqnarray}
 S_{0001}&=&{1\over 4\pi^{2}}\int_{0}^{\infty}d\omega\
 \Bigg\{ -{2r^{2}\over f}
  [(\omega-eA_{\tau})^{2}+m^{2}f+{f\over 4r^{2}}]^{1/2}  \nonumber\\
 &&+{1\over [(\omega-eA_{\tau})^{2}+m^{2}f+{f\over 4r^{2}}]^{1/2}}  \nonumber\\
 &&+{f\over 30r^{2}[(\omega-eA_{\tau})^{2}+m^{2}f+{f\over 4r^{2}}]^{3/2}} 
  \nonumber\\
 &&+{f^{2}\over 105r^{4}[(\omega-eA_{\tau})^{2}+m^{2}f+{f\over 4r^{2}}]^{5/2}}
   +\ldots  \nonumber\\
 &&-\left(
  -{2r^{2}\over f}\omega
  +\left({1\over 12}-m^{2}r^{2}
  \right){1\over \omega}+{2er^{2}A_{\tau}\over f}\right) \Bigg\}.
\end{eqnarray}
Performing the integrals over $\omega$ and expanding in the large mass 
limit yields
\begin{eqnarray}
 S_{0001}&\approx&{m^{2}r^{2}\over 8\pi^{2}}\left[
 -1+\ln\left({m^{2}f\over 4\lambda^{2}}\right)\right]
 -{mr^{2}eA_{\tau}\over 2\pi^{2}f^{1\over 2}} \nonumber\\
 &&-{1\over 96\pi^{2}}\ln\left({m^{2}f\over 4\lambda^{2}}\right)
 -{r^{2}(eA_{\tau})^{2}\over 4\pi^{2}f} \nonumber\\
 &&+{1\over m}\left({eA_{\tau}\over 48\pi^{2}f^{1\over 2}}-
  {r^{2}(eA_{\tau})^{3}\over 12\pi^{2}f^{3\over 2}}\right)
 +{7\over 3840m^{2}\pi^{2}r^{2}}  \nonumber\\
 &&+{1\over m^{3}}\left(
  {7eA_{\tau}\over 3840\pi^{2}r^{2}f^{1\over 2}}
  -{(eA_{\tau})^{3}\over 288\pi^{2}f^{3\over 2}}
  +{r^{2}(eA_{\tau})^{5}\over 80\pi^{2}f^{5\over 2}}\right)  \nonumber\\
 &&+{31\over 64512m^{4}\pi^{2}r^{4}}+\vartheta(m^{-5}).
\label{s0001expansion}\end{eqnarray}
In Eq.(\ref{s0001expansion}), the infrared cutoff parameter $\lambda$ 
has been introduced due to the lower limit $\omega=0$ on the integrals 
over $\omega$ \cite{bd,ahs}.  Comparison of Eq.(\ref{s0001expansion}) 
with Eq.(G2) of Ref.  \cite{ahs} shows that the presence of the gauge 
field has brought odd powers of $m$ into the asymptotic expansions for 
the $S_{ihjk}$.  Fortunately, when the large mass expansions of all of 
the $S_{ihjk}$ are computed, many are found to be of such a high order 
in inverse powers of $m$ that they do not contribute to the final 
expression for $G_{E,ren}$ since the DeWitt-Schwinger expansions are 
truncated at $\vartheta(m^{-2})$.  Note that the gauge field has also 
contributed terms of $\vartheta(m^{1})$ and $\vartheta(m^{0})$ to 
$S_{0001}$, as it does for some of the other $S_{ihjk}$.  In this 
work, only terms which contain negative powers of $m$ will be 
considered in the final DeWitt-Schwinger approximation.  Finally, all 
of the $S_{ihjk}$ are substituted into the general expression for 
$G_{E,ren}$
\begin{equation}
 G_{E,ren}=\{F(r)S_{ihjk}\},
\end{equation}
where the curly brackets $\{\ldots\}$ indicate the great number of 
terms to be added together, and like powers of $m$ are collected.

\section{Results}
\label{sec:results}

When all the substitutions are made and terms of like powers of $m$ 
are collected, the result is that the coefficients of the negative 
powers of $m$ constitute the DeWitt-Schwinger approximation to 
$G_{E,ren}$ for a complex scalar field in a static spherically 
symmetric spacetime.  A complex scalar field was chosen in order to 
make a connection to the earlier work of DeWitt \cite{dtgf,bsd} and 
Christensen \cite{chr} involving a real scalar field since the results 
of the present work must reduce to their results when the charge of 
the field vanishes.  In their work, the generalized DeWitt-Schwinger 
approximation for the Hadamard elementary function for a scalar field 
is
\begin{equation}
 {1\over 2}G^{(1)}=\langle\phi^{2}\rangle \approx{[a_{2}]\over 
 16\pi^{2}m^{2}},
\label{g1phi2ds}\end{equation}
where $[a_{2}]$ is given by \cite{dtgf,rhwh}
\begin{eqnarray}
 [a_{2}]&=&-{1\over 180}R^{\alpha\beta}R_{\alpha\beta}
  +{1\over 180}R^{\alpha\beta\gamma\delta}R_{\alpha\beta\gamma\delta}
  +{1\over 6}({1\over 5}-\xi)R_{;\alpha}{}^{\alpha}
  \nonumber\\
&&+{1\over 2}({1\over 6}-\xi)^2 R^2 -
   {e^2\over 12}F^{\alpha\beta}F_{\alpha\beta}.
\label{a2c0}\end{eqnarray} 
In the present work, the $m^{-2}$ term in the DeWitt-Schwinger 
approximation for $G_{E,ren}={1\over 2}G^{(1)}=\langle\phi^{2}\rangle$ 
is given in Euclidean space by
\begin{eqnarray}
 G_{E,m^{-2}}&=&{1\over m^{2}\,{{\pi }^2}}\Bigg(
 {\frac{1}{240\,{r^4}}} - 
 {\frac{1}{240\,{r^4}\,{{h}^2}}} + 
 {\frac{{\xi^2}\,{{R}^2}}{32}} \nonumber\\
 &&- 
 {\frac{{e^2}\,{{A_{\tau}'}^2}}
   {96\,f\,h}} + 
 {\frac{f'}{96\,{r^3}\,f\,{{h}^2}}} - 
 {\frac{f'}{96\,{r^3}\,f\,h}} + 
 {\frac{23\,{{f'}^2}}
   {5760\,{r^2}\,{{f}^2}\,{{h}^2}}} \nonumber\\
 &&+ 
 {\frac{{{f'}^2}}{576\,{r^2}\,{{f}^2}\,h}} - 
 {\frac{23\,{{f'}^3}}
   {1920\,r\,{{f}^3}\,{{h}^2}}} + 
 {\frac{7\,{{f'}^4}}
   {1280\,{{f}^4}\,{{h}^2}}} - 
 {\frac{3\,h'}{160\,{r^3}\,{{h}^3}}} \nonumber\\
 &&+ 
 {\frac{h'}{96\,{r^3}\,{{h}^2}}} - 
 {\frac{f'\,h'}{96\,{r^2}\,f\,{{h}^3}}} + 
 {\frac{f'\,h'}{576\,{r^2}\,f\,{{h}^2}}} -
 {\frac{{{f'}^2}\,h'}
   {80\,r\,{{f}^2}\,{{h}^3}}} \nonumber\\
 &&+ 
 {\frac{13\,{{f'}^3}\,h'}
   {1920\,{{f}^3}\,{{h}^3}}} - 
 {\frac{3\,{{h'}^2}}{640\,{r^2}\,{{h}^4}}} - 
 {\frac{11\,f'\,{{h'}^2}}
   {640\,r\,f\,{{h}^4}}} + 
 {\frac{5\,{{f'}^2}\,{{h'}^2}}
   {768\,{{f}^2}\,{{h}^4}}} \nonumber\\
 &&+ 
 {\frac{7\,{{h'}^3}}{240\,r\,{{h}^5}}} + 
 {\frac{7\,f'\,{{h'}^3}}
   {960\,f\,{{h}^5}}} + 
 {\frac{f''}{288\,{r^2}\,f\,{{h}^2}}} - 
 {\frac{f''}{288\,{r^2}\,f\,h}} \nonumber\\
 &&+ 
 {\frac{7\,f'\,f''}
   {320\,r\,{{f}^2}\,{{h}^2}}} - 
 {\frac{13\,{{f'}^2}\,f''}
   {960\,{{f}^3}\,{{h}^2}}} + 
 {\frac{13\,h'\,f''}
   {960\,r\,f\,{{h}^3}}} - 
 {\frac{5\,f'\,h'\,f''}
   {384\,{{f}^2}\,{{h}^3}}} \nonumber\\
 &&- 
 {\frac{19\,{{h'}^2}\,f''}
   {1920\,f\,{{h}^4}}} + 
 {\frac{{{f''}^2}}{192\,{{f}^2}\,{{h}^2}}} + 
 {\frac{h''}{240\,{r^2}\,{{h}^3}}} + 
 {\frac{f'\,h''}{120\,r\,f\,{{h}^3}}} \nonumber\\
 &&- 
 {\frac{{{f'}^2}\,h''}
   {384\,{{f}^2}\,{{h}^3}}} - 
 {\frac{13\,h'\,h''}{480\,r\,{{h}^4}}} - 
 {\frac{13\,f'\,h'\,h''}
   {1920\,f\,{{h}^4}}} + 
 {\frac{f''\,h''}{240\,f\,{{h}^3}}} \nonumber\\
 &&+ 
  \xi\,\left\{ -{\frac{{{R}^2}}{96}} + 
    \left( -{\frac{1}{48\,r\,h}} - 
       {\frac{f'}{192\,f\,h}} + 
       {\frac{h'}{192\,{{h}^2}}} \right) \,R'
      - {\frac{R''}{96\,h}} \right\}  \nonumber\\
 &&- 
 {\frac{f'''}{120\,r\,f\,{{h}^2}}} + 
 {\frac{f'\,f'''}
   {192\,{{f}^2}\,{{h}^2}}} + 
 {\frac{h'\,f'''}{160\,f\,{{h}^3}}} + 
 {\frac{h'''}{240\,r\,{{h}^3}}} \nonumber\\
 &&+ 
 {\frac{f'\,h'''}{960\,f\,{{h}^3}}} - 
 {\frac{f''''}{480\,f\,{{h}^2}}} \Bigg)
\label{gemm2}\end{eqnarray}
where the primes indicate differentiation with respect to $r$.  This 
is exactly the same result that is obtained by writing 
Eqs.(\ref{g1phi2ds})-(\ref{a2c0}) in terms of the metric functions of 
Eq.(\ref{eucmetric}).  Since the gauge field only appears here in the 
form $(eA_{\tau}')^{2}$, rotating back to Lorentzian space to obtain 
$G^{(1)}$ simply results in the appearance of a negative sign for this 
term.  This demonstrates the correspondence between the present method 
in a static spherically symmetric spacetime and the method of the 
generalized DeWitt-Schwinger expansion; they yield the same 
information for the real part of $G^{(1)}$ in the limit that the mass 
of the quantum field is large when compared to the inverse of the 
radius of curvature of the spacetime.

With the information in Eq.(\ref{gemm2}) previously known and 
accepted, it is fortunate that the present method not only duplicates 
that information but also yields more information.  The 
DeWitt-Schwinger point-splitting expansion is fully covariant and 
yields complete information about each of the coefficients in the 
expansion, indicating that all of the terms in a large mass 
DeWitt-Schwinger expansion proportional to inverse, even powers of $m$ 
have been determined by the generalized expansion.  These terms are 
constructed out of real curvature and electromagnetic field tensors 
and thus are incapable of yielding the imaginary part of $G^{(1)}$.  
Yet the definition of $G^{(1)}$ for a complex scalar field,
\begin{equation}
 G^{(1)}(x,x')\equiv \langle 0|\{\phi(x),\phi^{*}(x')\}|0\rangle,
\label{g1complexdef2}\end{equation}
shows that it is, in general, a complex quantity.  Any method for 
calculating $G^{(1)}$ that does not yield both real and imaginary 
pieces is incomplete in its determination of $G^{(1)}$.  DeWitt 
emphasized this limitation existed for the point-splitting procedure 
by explicitly stating that the expansion in inverse powers of $m^{2}$ 
was incapable of yielding the imaginary part of $G^{(1)}$ \cite{bsd}.  
Yet this imaginary part is non-vanishing even in flat space with a 
constant gauge field.  Classical field theory calculations of the 
current due to a complex scalar field under these conditions is not 
identically zero \cite{lhr}.  Schwinger's work \cite{schwinger} 
incorporated these flat space, constant field conditions specifically 
with the result being the non-vanishing Schwinger pair creation rate 
of Eq.(\ref{schwingerrate}).

The only way in which the imaginary part of $G^{(1)}$ may be obtained 
in a DeWitt-Schwinger expansion is by obtaining the terms proportional 
to inverse, odd powers of $m$.  It is possible to anticipate the form 
of the terms in the expansion for $G^{(1)}$.  Moving to units where 
$c=\hbar=1$ but $G\neq 1$, then $G^{(1)}$ and curvature tensors such 
as $R_{\mu\nu}$ are second order quantities with units of 
$(length)^{-2}$ while $m$ and $A_{\tau}$ have units of 
$(length)^{-1}$.  Using such power counting, Davies {\it et al.} 
\cite{dfcb} constructed the stress energy tensor of a conformally 
invariant scalar field in a conformally invariant spacetime.  In the 
present case of the expansion for $G^{(1)}$ the same procedure may be 
used.  The term proportional to $m^{-1}$ must be proportional to 
$(length)^{-3}$ in order for $G^{(1)}$ to remain second order.  Since 
Eq.(\ref{g1complexdef2}) holds in both flat and curved spacetimes, the 
present method should yield a non-vanishing imaginary part when 
$f(r)=h(r)=1$ and when the gauge field $A_{\tau}=\hbox{constant}$.  
Under these conditions, the $m^{-1}$ term of $G_{E}$ should be 
proportional to $(eA_{\tau})^{3}$ and $(eA_{\tau})r^{-2}$.  This power 
counting may be expanded by allowing the gauge field to be a function 
of $r$, with the result that additional terms such as 
$(eA_{\tau}')r^{-1}$ should appear in the expansion.

The $m^{-1}$ term in the DeWitt-Schwinger approximation for 
$G_{E,ren}$ in the present work is given in Euclidean space by
\begin{eqnarray}
 G_{E,m^{-1}}&=&{1\over m\,{{\pi }^2}}\Bigg(
 -{\frac{{(e\,A_{\tau})^3} }
    {24\,{{f}^{{\frac{3}{2}}}}}} + 
  {\frac{e\,A_{\tau}}
    {24\,{r^2}\,{{f}^{{\frac{1}{2}}}}}} - 
  {\frac{e\,R\,\xi\,A_{\tau}}
    {8\,{{f}^{{\frac{1}{2}}}}}} \nonumber\\
 &&- 
  {\frac{e\,A_{\tau}}
    {24\,{r^2}\,{{f}^{{\frac{1}{2}}}}\,h}} + 
  {\frac{e\,A_{\tau}'}
    {24\,r\,{{f}^{{\frac{1}{2}}}}\,h}} - 
  {\frac{e\,A_{\tau}\,f'}
    {16\,r\,{{f}^{{\frac{3}{2}}}}\,h}} - 
  {\frac{e\,A_{\tau}'\,f'}
    {48\,{{f}^{{\frac{3}{2}}}}\,h}} \nonumber\\
 &&+ 
  {\frac{3\,e\,A_{\tau}\,{{f'}^2}}
    {128\,{{f}^{{\frac{5}{2}}}}\,h}} + 
  {\frac{e\,A_{\tau}\,h'}
    {24\,r\,{{f}^{{\frac{1}{2}}}}\,{{h}^2}}} - 
  {\frac{e\,A_{\tau}'\,h'}
    {96\,{{f}^{{\frac{1}{2}}}}\,{{h}^2}}} + 
  {\frac{e\,A_{\tau}\,f'\,h'}
    {64\,{{f}^{{\frac{3}{2}}}}\,{{h}^2}}} \nonumber\\
 &&+ 
  {\frac{e\,A_{\tau}''}
    {48\,{{f}^{{\frac{1}{2}}}}\,h}} - 
  {\frac{e\,A_{\tau}\,f''}
    {32\,{{f}^{{\frac{3}{2}}}}\,h}} \Bigg).
\label{gemm1}\end{eqnarray}
All of these terms are proportional to odd powers of $A_{\tau}$ and 
its derivatives and, due to Eq.(\ref{AtauAt}), are thus all imaginary 
when rotated back to Lorentzian space.  All of these terms are also 
proportional to odd powers of the charge $e$ of the quantum field.  
Upon imposing charge conjugation symmetry on 
$\langle\phi^{2}\rangle={1\over 2}G^{(1)}$ \cite{ykluger}, these terms 
will not contribute to the calculation of the vacuum polarization in 
the spacetime.

It is remarkable that the two methods yield exactly the same result 
for the $m^{-2}$ term in their respective DeWitt-Schwinger expansions.  
As pointed out by Gibbons \cite{gwg}, while the DeWitt-Schwinger 
point-splitting expansion is widely used in renormalization theory, 
there is still much that is unknown about the type of series expansion 
that it really is.  The nonzero results of Eq.(\ref{gemm1}) highlight 
the major difference between the present method and the generalized 
DeWitt-Schwinger point-splitting expansion.  A major limitation of the 
DeWitt-Schwinger point-splitting expansion is that it is an asymptotic 
expansion limited to inverse powers of $m^{2}$.  This dictates that it 
is restricted to the calculation of only the real part of $G^{(1)}$ 
which, as stated above, is generally complex.  Yet the 
DeWitt-Schwinger expansion has the distinct advantage of being a 
completely general expansion with its pieces constructed from 
curvature and electromagnetic tensors.  Thus, once the point-splitting 
calculations have been performed and renormalization achieved by 
discarding terms proportional to nonnegative powers of $m$, the 
remainder is a completely general expression which may then be used to 
study any spacetime for which a metric exists.

The main limitation of the present method is that it is restricted to 
spacetimes for which the basis functions of the quantum field can be 
put into the WKB form like that of Eq.(\ref{wkb}).  The wave equation 
must therefore be separable.  Since this has been achieved for the 
Kerr metric \cite{bc}, the present work could conceivably be extended 
to Kerr spacetimes.

The major advantage of the present method over the DeWitt-Schwinger 
point-splitting expansion is that it can yield the imaginary part of 
$G^{(1)}$.  The calculation of the imaginary part of $G^{(1)}$ is 
important due to the definition of the current $\langle 
j^{\mu}\rangle$ in Eq.(\ref{divcurrent}), which is non-vanishing even 
in flat space with a constant gauge field.  The factor of $i$ in front 
of the expression, along with the subtraction of the complex conjugate 
of the derivatives of $G^{(1)}$, show that only the imaginary part of 
$G^{(1)}$ will contribute to the current.  The factor of $e$ in front 
of the expression for $\langle j^{\mu}\rangle$ requires the presence 
of the odd powers of $e$ in the imaginary part of $G^{(1)}$ if the 
current is to be constructed from $G^{(1)}$.  The final expression for 
the current must be proportional to even powers of $e$ if the current 
is not to vanish under charge conjugation symmetry \cite{ykluger}.  
Work is in progress to calculate the DeWitt-Schwinger approximation 
for the current in two ways; first, directly from Eqs.(\ref{gemm1}) 
and (\ref{divcurrent}), and second, starting from 
Eq.(\ref{divcurrent}) and proceeding through the steps outlined in 
this paper to calculate the DeWitt-Schwinger approximations for 
quantities such as $G^{(1)|\mu}$ that are required to construct 
$\langle j^{\mu}\rangle$.

Much of the previous research in quantum field theory in curved 
spacetimes dealt with the production of virtual particles due to 
vacuum polarization.  Whenever real particle production due to an 
electromagnetic field was to be studied, investigators most often used 
the pair creation rate derived by Schwinger for flat spacetime 
\cite{flatrate}.  While Schwinger's work assumed a constant electric 
field $E$, the authors in Ref.  \cite{flatrate} replaced that field 
with the spherically symmetric radial field of the 
Reissner-Nordstr\"{o}m spacetime.  With the curved space results in 
Eq.(\ref{gemm1}), it should now be possible to determine how the 
effects of curved space will cause the rate of charged particle 
production to differ from that of the flat space Schwinger rate.

\acknowledgements 

The author would like to thank William A. Hiscock and Paul R. Anderson 
for helpful discussions concerning this work.  This work was supported 
by a Radford University Seed Grant.



\begin{references}
\bibitem[*]{her} electronic mail address: rherman@runet.edu

\bibitem{mtw} C.\ W.\ Misner, K.\ S.\ Thorne, and J.\ A.\ Wheeler, 
  {\it Gravitation} (Freeman, San Francisco, 1973).

\bibitem{bd} N.\ D.\ Birrell and P.\ C.\ W.\ Davies, {\it Quantum fields 
 in curved space} (Cambridge University Press, Cambridge, England, 
 1982).

\bibitem{Schweber} S.\ S.\ Schweber, {\it An Introduction to Relativistic 
  Quantum Field Theory} (Row, Peterson, and Company, New York, 1961). 
  See chapter 16 for a discussion of the degree of divergences in QED.

\bibitem{BD1} J.\ D.\ Bjorken and S.\ D.\ Drell, {\it Relativistic 
  Quantum Mechanics} (McGraw-Hill, New York, 1964).

\bibitem{BD2} J.\ D.\ Bjorken and S.\ D.\ Drell, {\it Relativistic 
 Quantum Fields} (McGraw-Hill, New York, 1965).

\bibitem{bsd} B.\ S.\ DeWitt, Phys. Rep. {\bf 19C}(1), 297 (1975).

\bibitem{schwinger} J.\ S.\ Schwinger, Phys.\ Rev.\ {\bf 82} 664 (1951).

\bibitem{schwinger6} J.\ S.\ Schwinger, Phys.\ Rev.\ {\bf 82} 664 (1951).
 Section VI contains the relevant perturbation theory.

\bibitem{dtgf} B.\ S.\ DeWitt, {\it The Dynamical Theory of Groups and 
  Fields} (Gordon and Breach, New York, 1965).

\bibitem{chr} S.\ M.\ Christensen, Phys.\ Rev.\ D {\bf 14}, 2490 (1976).

\bibitem{chrthesis} S.\ M.\ Christensen, Ph.\ D.\ thesis, University 
 of Texas at Austin, 1975.

\bibitem{pc} P.\ Candelas, Phys.\ Rev.\ D {\bf 21}, 2185 (1980).

\bibitem{vpf1} V.\ P.\ Frolov, Phys.\ Rev.\ D {\bf 26}, 
 954 (1982).

\bibitem{vpf2} V.\ P.\ Frolov, in {\it Quantum Gravity}, proceedings 
 of the Second Seminar on Quantum Gravity, Moscow, USSR, edited by 
 M.\ Markov and P.\ C.\ West (Plenum, New York, 1984).

\bibitem{ch} P.\ Candelas and K.\ W.\ Howard, Phys.\ Rev.\ D {\bf 21}, 
 1618 (1984).

\bibitem{cj} P.\ Candelas and B.\ P.\ Jensen, Phys.\ Rev.\ D {\bf 33}, 
 1596 (1986).

\bibitem{pra} P.\ R.\ Anderson, Phys.\ Rev.\ D {\bf 41}, 1152 (1990).

\bibitem{hch} K.\ W.\ Howard and P.\ Candelas, Phys.\ Rev.\ Lett.\ {\bf 
 53}, 403 (1984); K.\ W.\ Howard, Phys.\ Rev.\ D {\bf 30}, 2532 
 (1984).

\bibitem{fz} V.\ P.\ Frolov and A.\ I.\ Zel'nikov, Phys.\ Rev.\ D {\bf 
 29}, 1057 (1984).
  
\bibitem{bop} M.\ R.\ Brown, A.\ C.\ Ottewill, and D.\ N.\ Page, Phys.\ 
 Rev.\ D {\bf 33}, 2840 (1986).

\bibitem {ahs} P.\ R.\ Anderson, W.\ A.\ Hiscock, and D.\ A.\ Samuel, 
 Phys.\ Rev.\ D {\bf 51}, 4337 (1995).

\bibitem{rhthesis} Rhett Herman, Ph.\ D.\ thesis, Montana State 
 University, 1996. 

\bibitem{rhwhunpub} R.\ Herman and W.\ A.\ Hiscock, unpublished.

\bibitem{rhwh} Rhett Herman and William A.\ Hiscock, Phys.\ Rev.\ D {\bf 
  53}, 3285 (1996).

\bibitem{ww}E.\ T.\ Whittaker and G.\ N.\ Watson, {\it A Course of 
 Modern Analysis} (Cambridge University Press, Cambridge, England, 
 1927).

\bibitem{lhr} L.\ H.\ Ryder, {\it Quantum Field Theory} (Cambridge 
 University Press, Cambridge, England, 1985).

\bibitem{dfcb} P.\ C.\ W.\ Davies, S.\ A.\ Fulling, S.\ M.\ Christensen,
  T.\ S.\ Bunch, Ann. Phys. {\bf 109}, 108 (1977).

\bibitem{ykluger} Y.\ Kluger, J.\ M.\ Eisenberg, B.\ Svetitsky, F.\ 
Cooper, and E.\ Mottola, Phys.\ Rev.\ D {\bf 45}, 4659 (1992).

\bibitem{gwg} G.\ W.\ Gibbons, in {\it General Relativity: An 
 Einstein Centenary Survey}, edited by S.\ W.\ Hawking and W.\ Israel 
 (Cambridge University Press, Cambridge, England, 1979).

\bibitem{bc} B.\ Carter, Phys.\ Rev.\ {\bf 174}, 1559 (1968).

\bibitem{flatrate} See, for example, I.\ D.\ Novikov and A.\ A.\ 
 Starobinski\`i, Sov.\ Phys.\ JETP {\bf 51}, 1 (1980), and W.\ A.\ 
 Hiscock and L.\ Weems, Phys.\ Rev.\ D {\bf 41}, 1142 (1990).
 
\end{references}
\end{document}